\documentclass[twocolumn,showpacs,preprintnumbers,amsmath,amssymb]{revtex4}
\usepackage{graphicx}
\begin{document}
\title{Transport through strongly correlated triple quantum dot}
\author{D. Krychowski, M. Antkiewicz, S. Lipi\'{n}ski}
\affiliation{%
Institute of Molecular Physics, Polish Academy of Sciences\\M. Smoluchowskiego 17,
60-179 Pozna\'{n}, Poland
}%
\date{\today}
\begin{abstract}
Strong electron correlations are discussed for the three capacitively coupled quantum dots, each of which is connected to a separate pair of electrodes. The finite-$U$ mean field slave boson approach is used. The analysis is carried out for both repulsive and attractive  intra- and inter-dot interactions. Depending on the ratio and the sign of interaction parameters and occupation, either charge ordered states or  different spin, spin-charge and charge Kondo resonances  arise.
\end{abstract}

\maketitle

\section{Introduction}
Due to the recent technical  improvements, three coupled   quantum dots (TQD) have attracted considerable interest both  for fundamental physics and for potential applications, in particular in the field of quantum information processing \cite{Gaudreau}. Experimentally, the triple quantum-dot systems have been realized in AlGaAs/GaAs heterostuctures \cite{Vidan}, self-assembled InAs \cite{Amaha}, and single wall carbon nanotubes \cite{Grove}. Due to the wide experimental tunability TQDs provide an ideal platform for studying  many-body states \cite{Wojcik}. The charge stability diagrams have been examined using standard detection techniques and there are reports on finding different degeneracy points \cite{Seo}.  Apart from spin also charge Kondo effects are predicted in TQD \cite{Yoo}. Nowadays, not only  repulsive interactions between electrons are studied, but also attractive, and they are also experimentally observed in quantum dot systems \cite{Hong}. Apart from bipolaronic, excitonic, plasmonic or chemical mechanisms usually invoked as a source of electron attraction, also  coupling to mechanical resonator has recently  been suggested as a tool for  engineering attractive interactions in quantum dots \cite{Sechenyi}. In the present paper we discuss transport through TQD in the strongly correlated regime for the cases of repulsive and attractive electron-electron interactions.

\section{Model and formalism}
We consider three capacitively coupled QDs in triangular arrangement with each of the dots contacted to its own source and drain electrodes (Fig. 1a). The system is modelled by three-impurity Anderson Hamiltonian:
\begin{eqnarray}
&&{\cal{H}}=\sum_{i\sigma}E_{d}n_{i\sigma}+\sum_{k\alpha i\sigma}E_{kl\alpha}n_{k\alpha i\sigma}+\\
&&\nonumber\sum_{k\alpha i\sigma}t(c^{\dagger}_{k\alpha i\sigma}d_{i\sigma}+h.c.)+U\sum_{i}n_{i\uparrow}n_{i\downarrow}+U'\sum_{ii'\sigma \sigma'}n_{i\sigma}n_{i'\sigma'},
\end{eqnarray}
where $c^{\dagger}_{k\alpha i\sigma}$ creates an electron state in the $i$-th ($i = 1,2,3$) left  or right electrode $\alpha = L(R)$.  $n_{i\sigma}= d^{\dagger}_{i\sigma}d_{i\sigma}$ is  the occupation operator of the dot $i$, $E_{ki\alpha}$ ($E_{d}$) denote energies of electrons in the lead (dot). Intra- and interdot Coulomb interactions are parametrized by $U$, $U'$. To analyze correlation effects we use finite $U$ slave boson mean field approach (SBMFA) of Kotliar and Ruckenstein (K-R) \cite{Kotliar} and introduce a set of boson operators, which project the state space onto subspaces of different occupation numbers. Boson e projects onto state with no electron at the dots, six  $p_{i\sigma}$ bosons project onto singly occupied states $|i\sigma\rangle$, three doubly occupied states at the same dot are represented by bosons $d_{i}$, twelve two-electron states with electrons occupying two  dots are generated by bosons $d_{ij\sigma\sigma'}$, twelve bosons $t_{i,j\sigma}$ represent three-electron states characterized by double and single occupancy of the two dots and eight  $t_{\sigma\sigma'\sigma''}$ opeators project onto triple occupied states with single electrons at the dots. Above half filling one can still label the states or their corresponding bosons by electron quantum numbers, but for brevity of notation it is more convenient to use hole quantum numbers. The four-electron (two-hole) states are then represented by auxiliary bosons $f$: $f_{i}$ – empty dot $i$ and $f_{ij\sigma\sigma'}$ – two holes at two dots.
The occupation by five electrons (single hole) is described by boson $q$ and full filling by boson $s$. Slave boson representation expands the space of states and in order  to eliminate unphysical states one has to  introduce additional constraints ensuring the completeness of states and charge conservations. This is achieved by supplementing  the effective slave boson Hamiltonian (2) by additional terms with Lagrange multipliers ($\lambda$, $\lambda_{i\sigma}$):
\begin{eqnarray}
&&\nonumber{\cal{\widetilde{H}}}=\sum_{i}E_{d}f^{\dagger}_{i\sigma}f_{i\sigma}+\sum_{k\alpha\sigma}E_{ki\alpha}c^{\dagger}_{ki\alpha\sigma}c_{ki\alpha\sigma}
+\\&&\sum_{k\alpha\sigma}t(c^{\dagger}_{ki\alpha\sigma}z_{i\sigma}f_{i\sigma}+h.c.)+U\sum_{i}d^{\dagger}_{i}d_{i}
+\\&&\nonumber U'\sum_{ij\sigma\sigma',i<j}d^{\dagger}_{ij\sigma\sigma'}d_{ij\sigma\sigma'}+
(U+2U')\sum_{ij\sigma,i\neq j}t^{\dagger}_{i,j\sigma}t_{i,j\sigma}+\\&&\nonumber 3U'\sum_{\sigma\sigma'\sigma''}t^{\dagger}_{\sigma\sigma'\sigma''}t_{\sigma\sigma'\sigma''}+(2U+4U')\sum_{i}f^{\dagger}_{i}f_{i}+
\\&&\nonumber (U+5U')\sum_{ij\sigma\sigma',i<j}f^{\dagger}_{ij\sigma\sigma'}f_{ij\sigma\sigma'}+(2U+8U')\sum_{i\sigma}q^{\dagger}_{i\sigma}q_{i\sigma}
\\&&\nonumber+(3U+12U')s^{\dagger}s+\lambda(I-1)+\sum_{i\sigma}\lambda_{i\sigma}(f^{\dagger}_{i\sigma}f_{i\sigma}-Q_{i\sigma}),
\end{eqnarray}
where $Q_{i\sigma}=p^{\dagger}_{i\sigma}p_{i\sigma}+d^{\dagger}_{i}d_{i}+\sum_{j\sigma'}d^{\dagger}_{ij\sigma\sigma'}d_{ij\sigma\sigma'}
+\sum_{j\sigma'}t^{\dagger}_{i,j\sigma'}t_{i,j\sigma'}+\sum_{j}t^{\dagger}_{j,i\sigma}t_{j,i\sigma}
+\sum_{\sigma\sigma'}t^{\dagger}_{\sigma\sigma'\sigma''}t_{\sigma\sigma'\sigma''}
+\sum_{j\sigma'}f^{\dagger}_{ij\overline{\sigma}\sigma'}f_{ij\overline{\sigma}\sigma'}
+\sum_{j}f^{\dagger}_{j}f_{j}+\sum_{i'<j,\sigma'\sigma''}f^{\dagger}_{i'j\sigma'\sigma''}f_{i'j\sigma'\sigma''}+
f^{\dagger}_{i'j\sigma'\sigma''}f_{i'j\sigma'\sigma''}+q^{\dagger}_{i\overline{\sigma}}q_{i\overline{\sigma}}
+\sum_{j\sigma'}q^{\dagger}_{j\sigma'}q_{j\sigma'}$,  $I=e^{\dagger}e+\sum_{i\sigma}p^{\dagger}_{i\sigma}p_{i\sigma}+\sum_{i}d^{\dagger}_{i}d_{i}
+\sum_{ij\sigma\sigma',i<j}d^{\dagger}_{ij\sigma\sigma'}d_{ij\sigma\sigma'}+\sum_{ij\sigma,i\neq j}t^{\dagger}_{i,j\sigma}t_{i,j\sigma}
+\sum_{\sigma\sigma'\sigma''}t^{\dagger}_{\sigma\sigma'\sigma''}t_{\sigma\sigma'\sigma''}+\sum_{i}f^{\dagger}_{i}f_{i}
+\sum_{ij\sigma\sigma',i<j}f^{\dagger}_{ij\sigma\sigma'}f_{ij\sigma\sigma'}
+\sum_{i\sigma}q^{\dagger}_{i\sigma}q_{i\sigma}+s^{\dagger}s$ are the conservation of charge and completeness relations and $z_{i\sigma}=(e^{\dagger}p_{i\sigma}+p^{\dagger}_{i\overline{\sigma}}d_{i}
+\sum_{j\sigma'}p^{\dagger}_{j\sigma'}d_{ij\sigma\sigma'}
+\sum_{j}d^{\dagger}_{j}t_{j,i\sigma}+\sum_{j\sigma'}d^{\dagger}_{ij\overline{\sigma}\sigma'}t_{i,j\sigma'}+
\sum_{i'<j,\sigma'\sigma''}d^{\dagger}_{i'j\sigma'\sigma''}t_{\sigma\sigma'\sigma''}
+\sum_{j}t^{\dagger}_{\overline{j},i\overline{\sigma}}f_{j}
+\sum_{j\sigma'}t^{\dagger}_{i,j\sigma'}f_{ij\overline{\sigma}\sigma'}
+\sum_{i'<j,\sigma'\sigma''}t^{\dagger}_{\overline{\sigma}\overline{\sigma'}\overline{\sigma''}}
f_{i'j\sigma'\sigma''}+f^{\dagger}_{i}q_{i\overline{\sigma}}
+\sum_{j\sigma'}f^{\dagger}_{ij\sigma\sigma'}q_{j\sigma'}+q^{\dagger}_{i\sigma}s)/\sqrt{1-Q_{i\sigma}}\sqrt{Q_{i\sigma}}$ renormalizes dot – lead hybridization. The form of renormalization factors $z_{i\sigma}$ expressed  in terms of the  roots of $Q_{i\sigma}$ is chosen in order to obtain the correct MFA limit in the uncorrelated case. K-R  formalism suffers from the lack of spin rotational  invariance and this may lead to incomplete conclusions about spin fluctuations. Another drawback of K-R approach  is an  overestimation od  Kondo temperature in the strongly correlated limit \cite{Lobos}. The mean-field solutions are found from the minimum of the free energy with respect to the mean values of  SB operators and Lagrange multipliers. In this approximation the problem of interacting electrons is formally reduced to the effective free-particle picture with the renormalized hopping integrals and renormalized dot energies.
\begin{figure}[b!]
\includegraphics[width=0.44\linewidth]{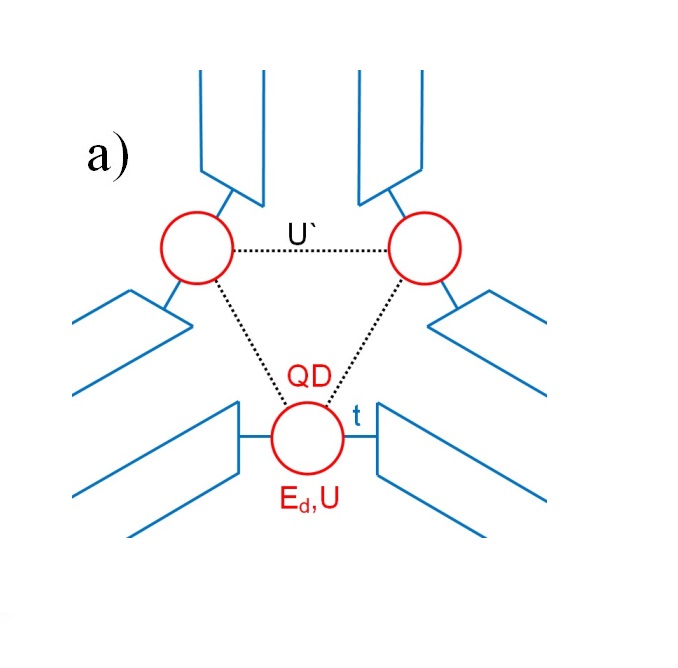}
\includegraphics[width=0.48\linewidth]{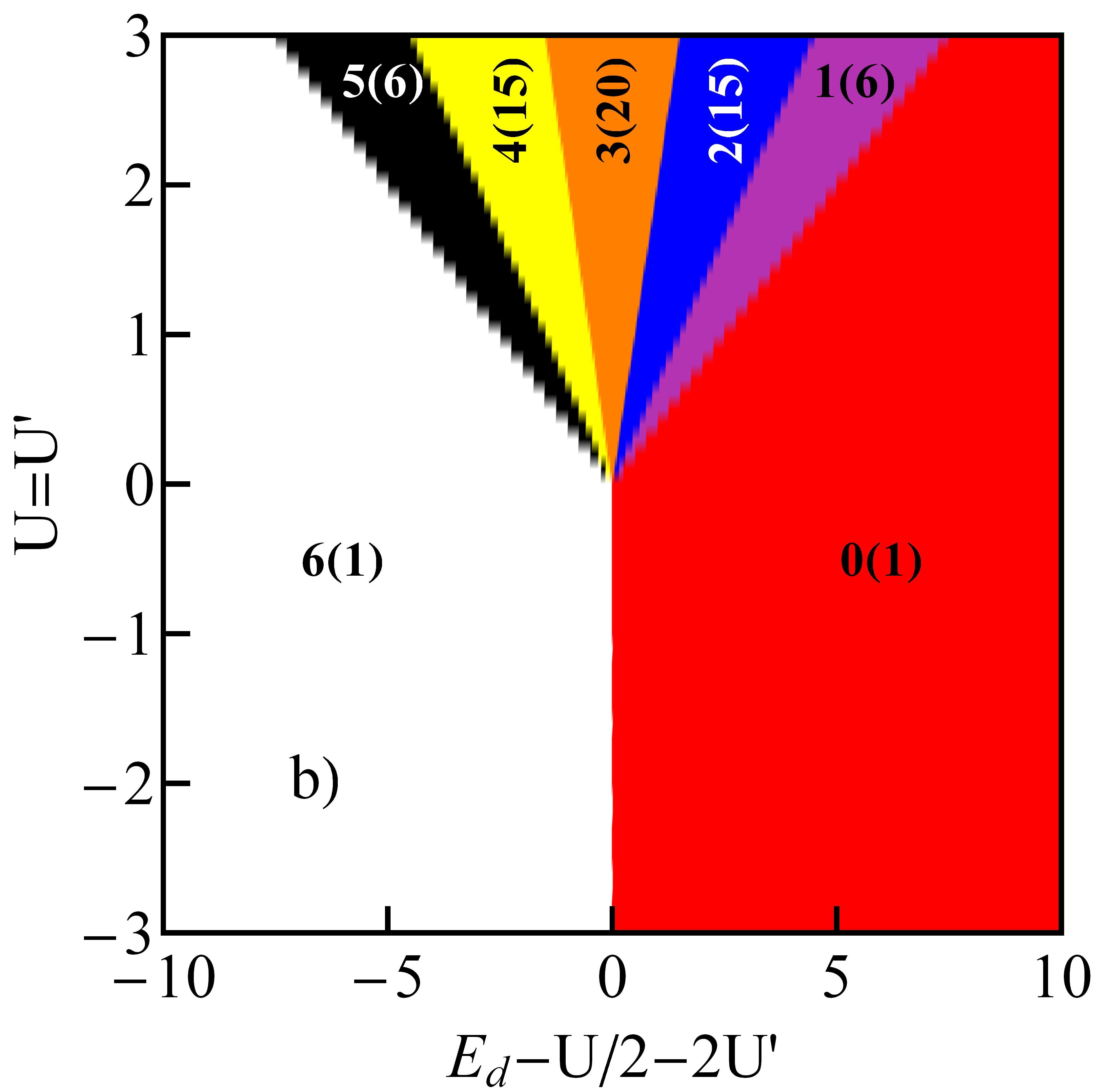}\\
\includegraphics[width=0.48\linewidth]{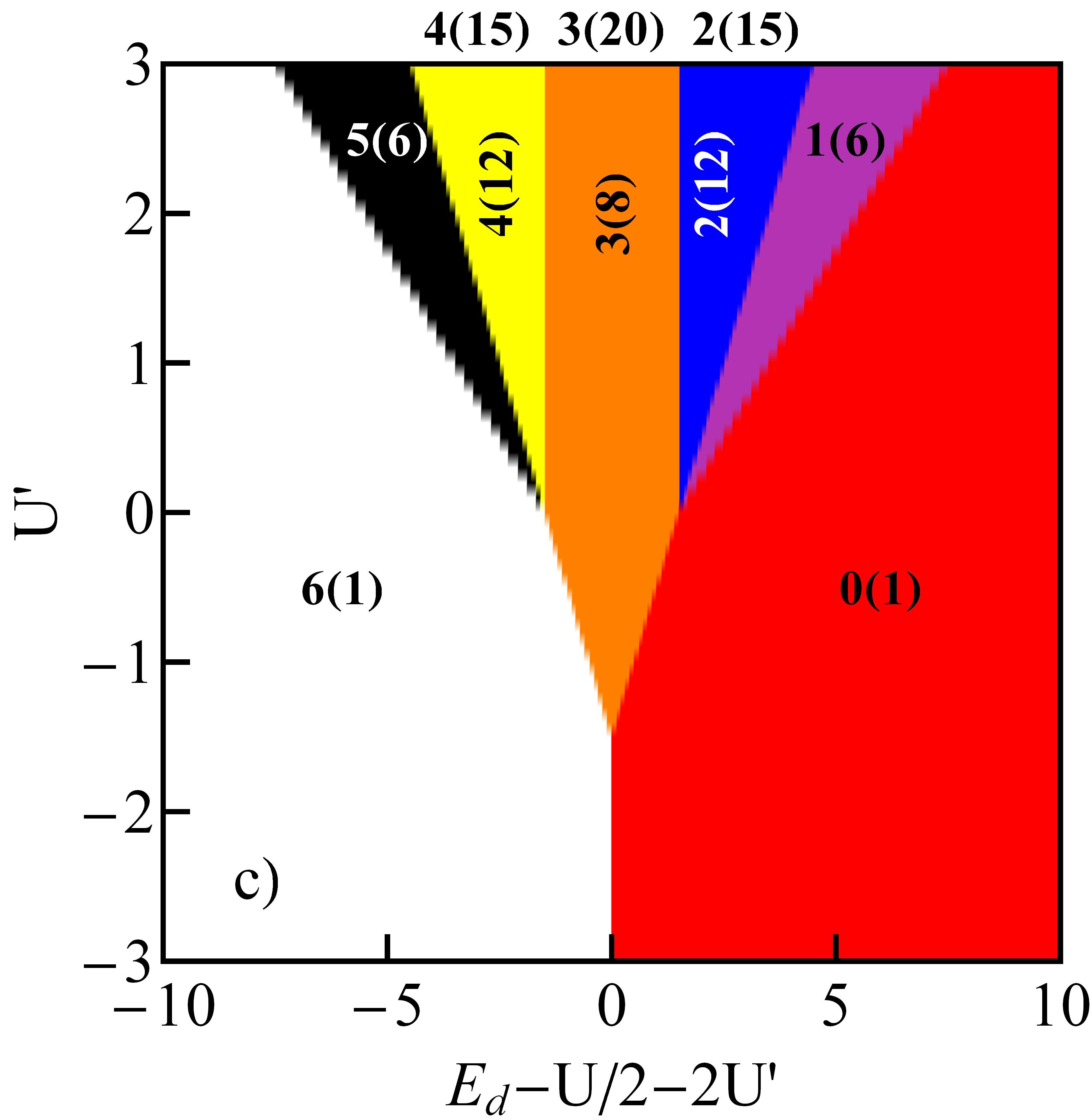}
\includegraphics[width=0.48\linewidth]{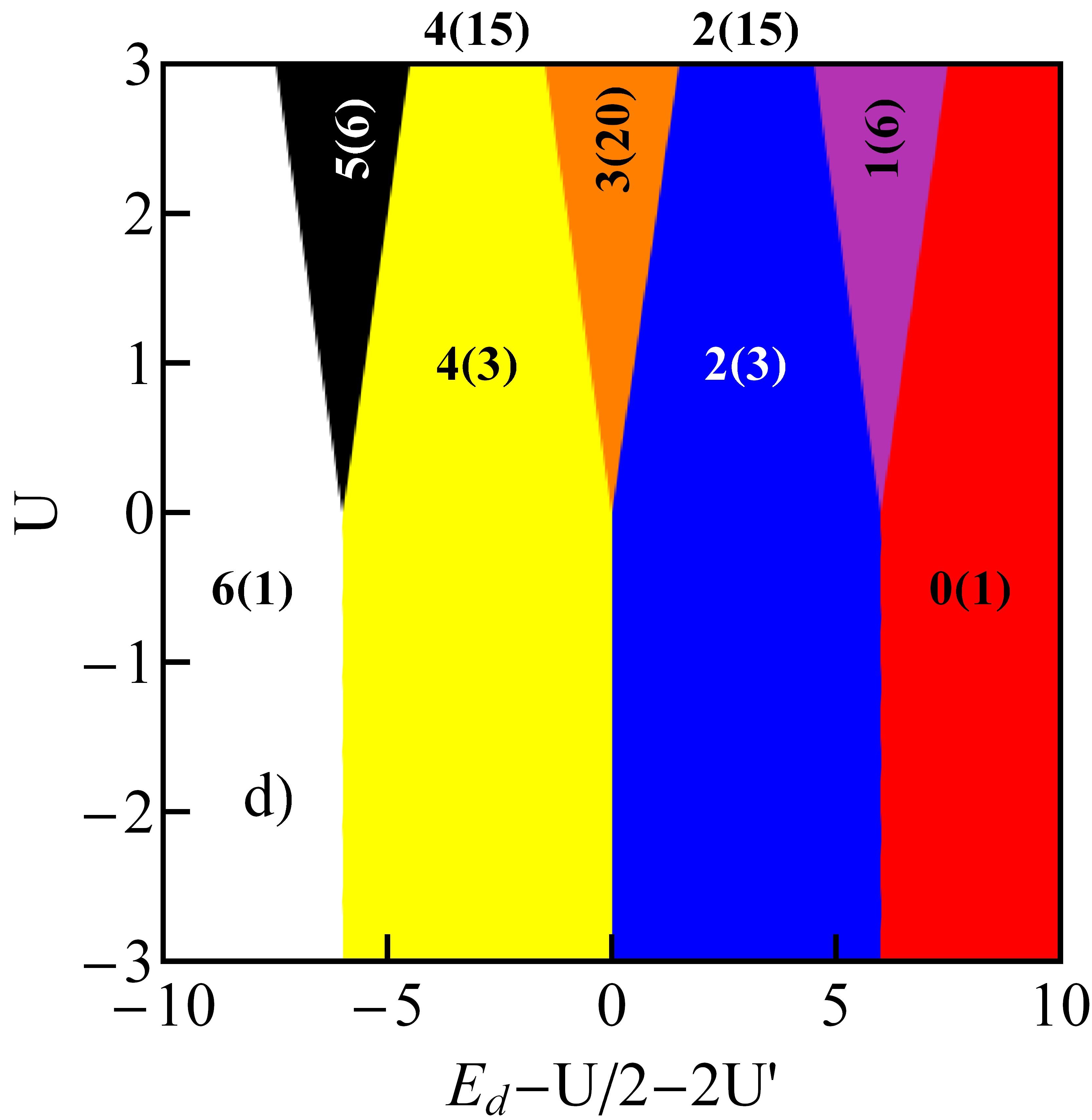}
\caption{\label{fig1} (Color online) 1(a) – Schematic view of TQD system, (b-d) Ground state diagrams of the isolated TQD system (atomic limit, $\Gamma = 0$) drawn as a function of site energy(gate voltage) and interaction parameters for the cases b)  $U = U'$, c) $U' < U$ ($U = 3$), d) $U < U'$ ($U' = 3$). The first number in a given charge region shows the total occupation, and the second, in parentheses informs about  degeneracy of the ground state.}
\end{figure}
The total conductance is a sum of single dot contributions, each of which can be measured separately ${\cal{G}} = \sum_{i}{\cal{G}}_{i}$.   ${\cal{G}}_{i}$  is given by Landauer type formula ${\cal{G}}_{i}=\frac{e^{2}}{h}\sum_{\sigma}\int(-\frac{df(E)}{dE}){\cal{T}}_{i\sigma}(E)dE$,
where ${\cal{T}}_{i\sigma}(E)=-Im[\widetilde{\Gamma}_{i\sigma}G^{R}_{i\sigma}(E)]$
and $f(E)$ is the Fermi distribution function, $G^{R}_{i\sigma}(E)=(E-\widetilde{E}_{i\sigma}+i\widetilde{\Gamma}_{i\sigma})^{-1}$ denotes the retarded Green’s function of the dot $i$, $\widetilde{E}_{i\sigma}=E_{d}+\lambda_{i\sigma}$ is the position of the Kondo resonance and $\widetilde{\Gamma}_{i\sigma}$ is the renormalized  coupling strength to the electrode of dot $i$, which for the assumed rectangular density of states $1/2D$ for $|E|<D$ is given by $\widetilde{\Gamma}_{i\sigma}=\frac{\pi t^{2}z^{2}_{i\sigma}}{2D}$. To follow the evolution of the system with the change of the strength of the interactions, we also study charge fluctuations $\Delta N^{2}=\langle N^{2}\rangle - \langle N\rangle^{2}$ ($N = \sum_{i\sigma} N_{i\sigma}$, $N_{i\sigma} = \langle f^{\dagger}_{i\sigma}f_{i\sigma}\rangle$), local moments $M = \langle(N_{\uparrow}-N_{\downarrow})^{2}\rangle^{1/2}$ and interdot fluctuations $\Delta T = \langle(N_{i} – N_{j})^{2}\rangle^{1/2}$.

\section{Results and discussion}
Throughout this paper we set $\hbar$=$k_{B}$=$|e|$=1 and we use relative energy units taking $D/50$ as the unit. Coupling of electrodes to the dots is assumed $\Gamma=0.05$.
Before we describe  transport properties of TQD system, we first present  ground state diagrams of the isolated interacting dots (atomic limit, $\Gamma = 0$) drawn as a function of gate voltage and interaction parameters for the  three  cases:   $U = U'$ (Fig. 1b), fixed $U$ and $U'\leq U$ (Fig. 1c) and fixed $U’$ and $U \leq U'$ (Fig. 1d).  Apart from showing charge stability regions also the degenerations of the  ground states corresponding to a given occupation are marked. Three-fold, four-fold and seven-fold charge degeneration points are visible and as can be seen, the same charge degeneracy can correspond to different degenerations of states. For $U' < U$ electrons  prefer occupation of different dots and for $U' > U$,  on the contrary, they tend to place at the same dot. For the fully symmetric TQD system ($U = U'$)   the slave bosons for a given occupation number are equal (Fig. 2b).
\begin{figure}[t!]
\includegraphics[width=0.48\linewidth]{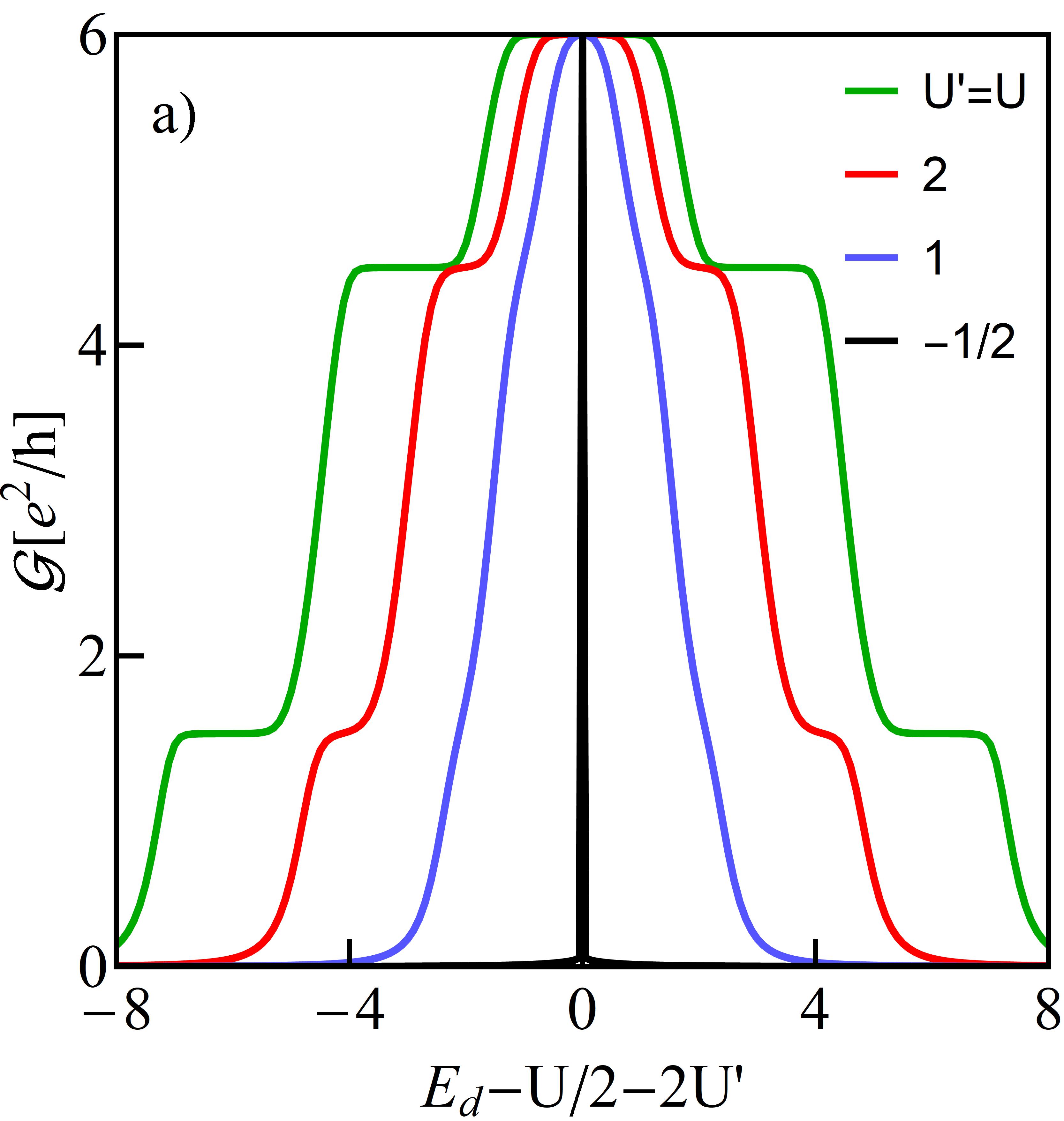}
\includegraphics[width=0.48\linewidth]{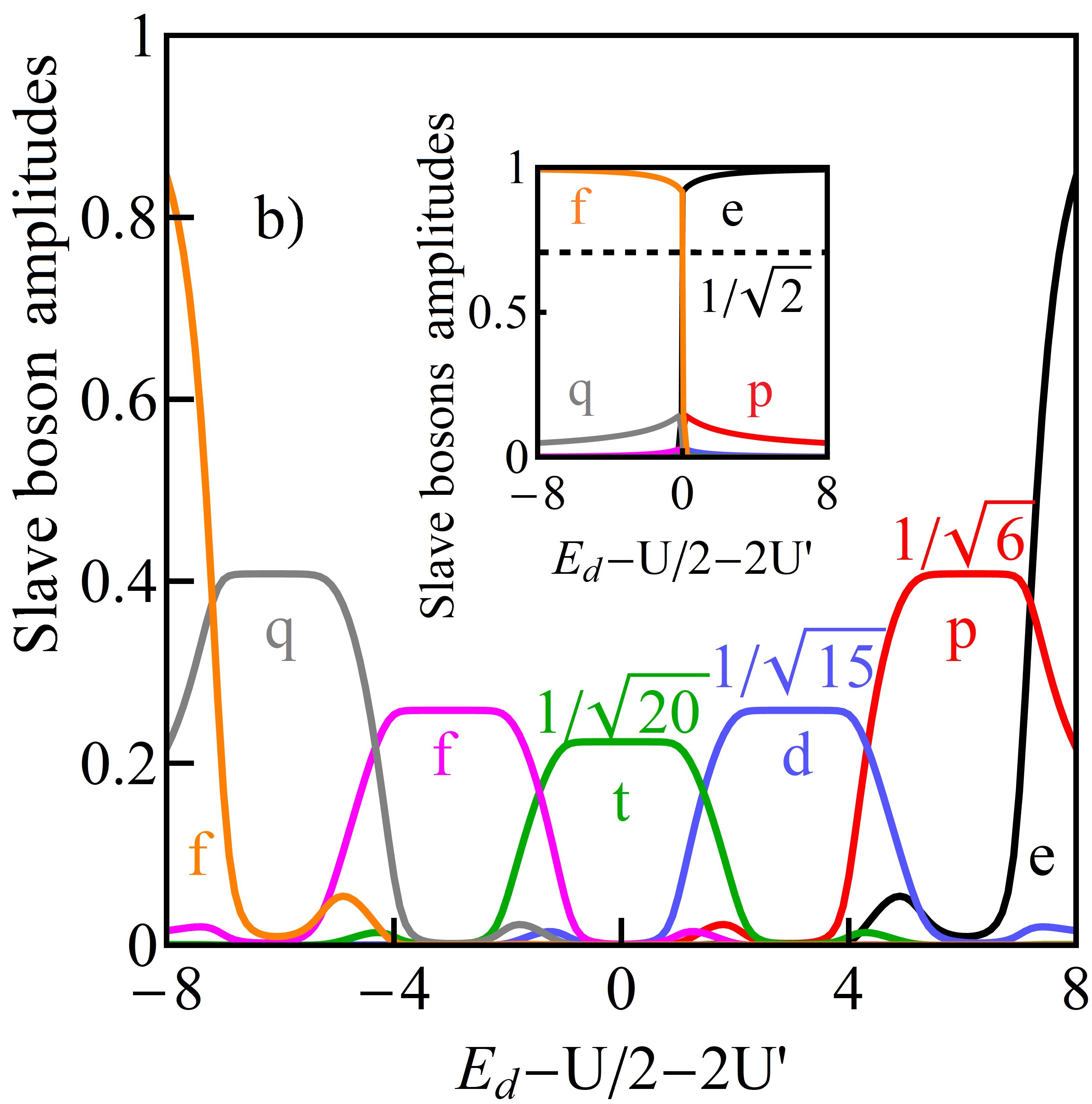}\\
\includegraphics[width=0.48\linewidth]{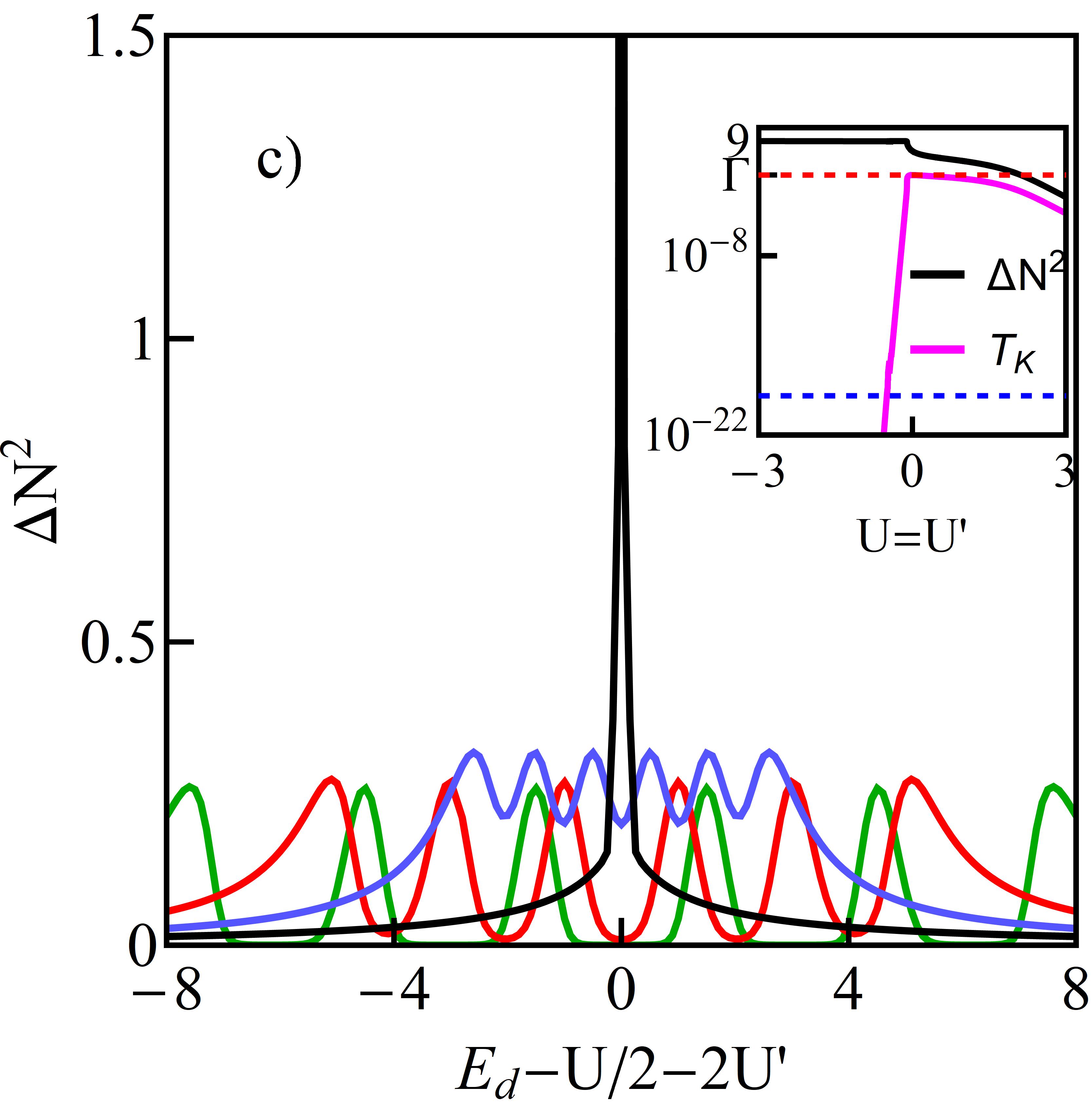}
\includegraphics[width=0.48\linewidth]{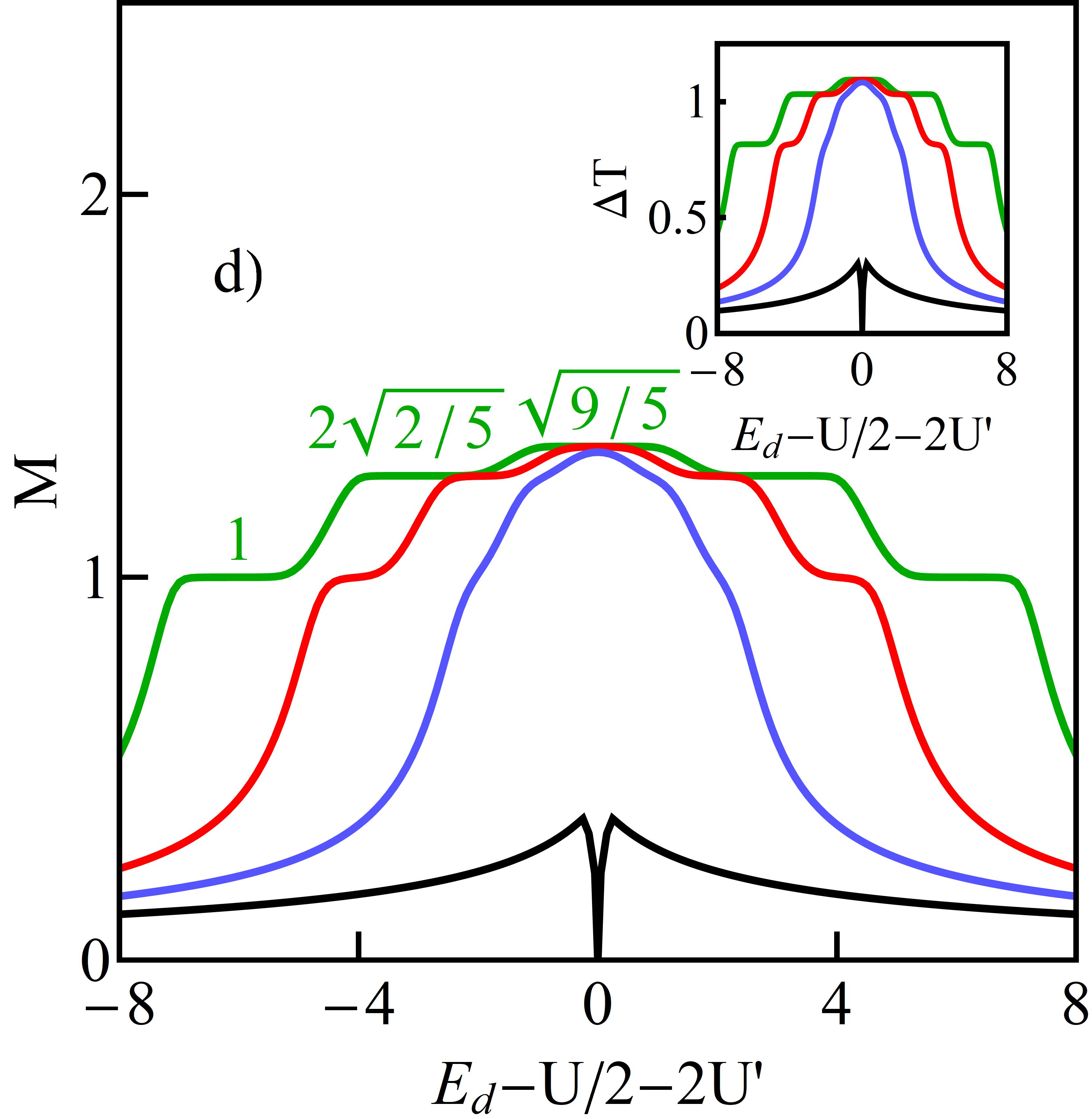}
\caption{\label{fig2} (Color online) a) Conductance vs. gate voltage for the selected  interaction parameters $U = U'$
 b) Slave boson amplitudes drawn for different site energies. The black dashed line in the inset marks values corresponding to charge Kondo state. The plotted numbers are the values of the respective SB amplitudes in the regions where they dominate. c) Gate dependencies of occupation fluctuations $\Delta N^{2}$ plotted for the same choice of interaction parameters as in Fig.2a. Inset presents $\Delta N^{2}$ and Kondo temperature vs. interaction parameter for e-h symmetry point, where dashed red and blue lines denote $\Gamma$ and $T_{K}$ for $U=0$ and $U=-1/2$. d) Gate dependencies of local moment $M$. The  plotted numbers are the magnetic moments in different occupation regions for $U = U’$. Inset shows interdot occupation fluctuation $\Delta T$ ($\Delta T\approx\sqrt{2/3}M$).}
\end{figure}
For strong interactions ($U = 3, 2$) gate dependencies of  conductances and occupations (not presented) are characterized by clear equal plateaus (Fig. 2a), which point on  the occurrence of  spin-orbital SU(6) Kondo effects. According to the Friedel sum rule  conductance  values   ${\cal{G}} = (e^{2}/h)\sum_{i\sigma}sin^{2}(\pi N_{i\sigma})$ are equal  ${\cal{G}} = 3/2(e^{2}/h)$ for  $N = 1$ and $5$, ${\cal{G}} = 9/2(e^{2}/h)$ ($N = 2, 4$) and  ${\cal{G}} = 6(e^{2}/h)$ ($N = 3$). Spins and interdot charge  polarization effectively fluctuate  in this state due to cotunneling processes. This  manifests in similar gate dependencies of  local moments  and interdot fluctuations presented on Fig. 2c. For low interaction values, charge fluctuations come into play and plateaus disappear ($U \sim 1$). For attractive interactions  $U' < 0$  the empty state and fully occupied state degenerate at electron – hole (e-h) symmetry point and hybridization generates tunneling within this degenerate manifold. The charge pseudospin flips from \textit{down} ($N = 0$) to \textit{up} ($N= 6$) state are accompanied by a coherent movement of six electrons into and out of the system of the  coupled interacting dots. These processes quench the charge pseudospin and nondegenerate charge Kondo (CK) state is formed. Transition amplitudes between  the mentioned degenerate charge states are of sixth order in $t$ and therefore the corresponding Kondo temperature is extremely small ($T_{K}\approx1.2\times10^{-19}$). For spin-orbital Kondo states charge fluctuations $\Delta N^{2}$  disappear and spin and interdot  charge polarization fluctuate ($M \neq 0$, $\Delta T \neq 0$). In charge Kondo state in turn, $\Delta N^{2}$ is maximal and  reaches value $9$  ($\Delta N^{2}\approx36(s^{2}-s^{4})$) and  $M$, $\Delta T$ vanish (Fig. 2c,d). Kondo temperature of CK state rapidly decreases with the increase of  $|U|$.

Fig. 3 illustrates evolution of the system with the decrease of interdot interaction. Apart from $N = 3$ region, the steps of conductance for other intermediate occupancies gradually disappear and only for $N = 3$ charge fluctuations remain totally suppressed. In other regions gradual transition to the mixed valence (MV) states is observed.
For $U' = 0$ only a single plateau for  $N = 3$ is seen and this case  corresponds to 3$\times$SU(2) Kondo effect - independent spin Kondo screening on each dot.
\begin{figure}[t!]
\includegraphics[width=0.48\linewidth]{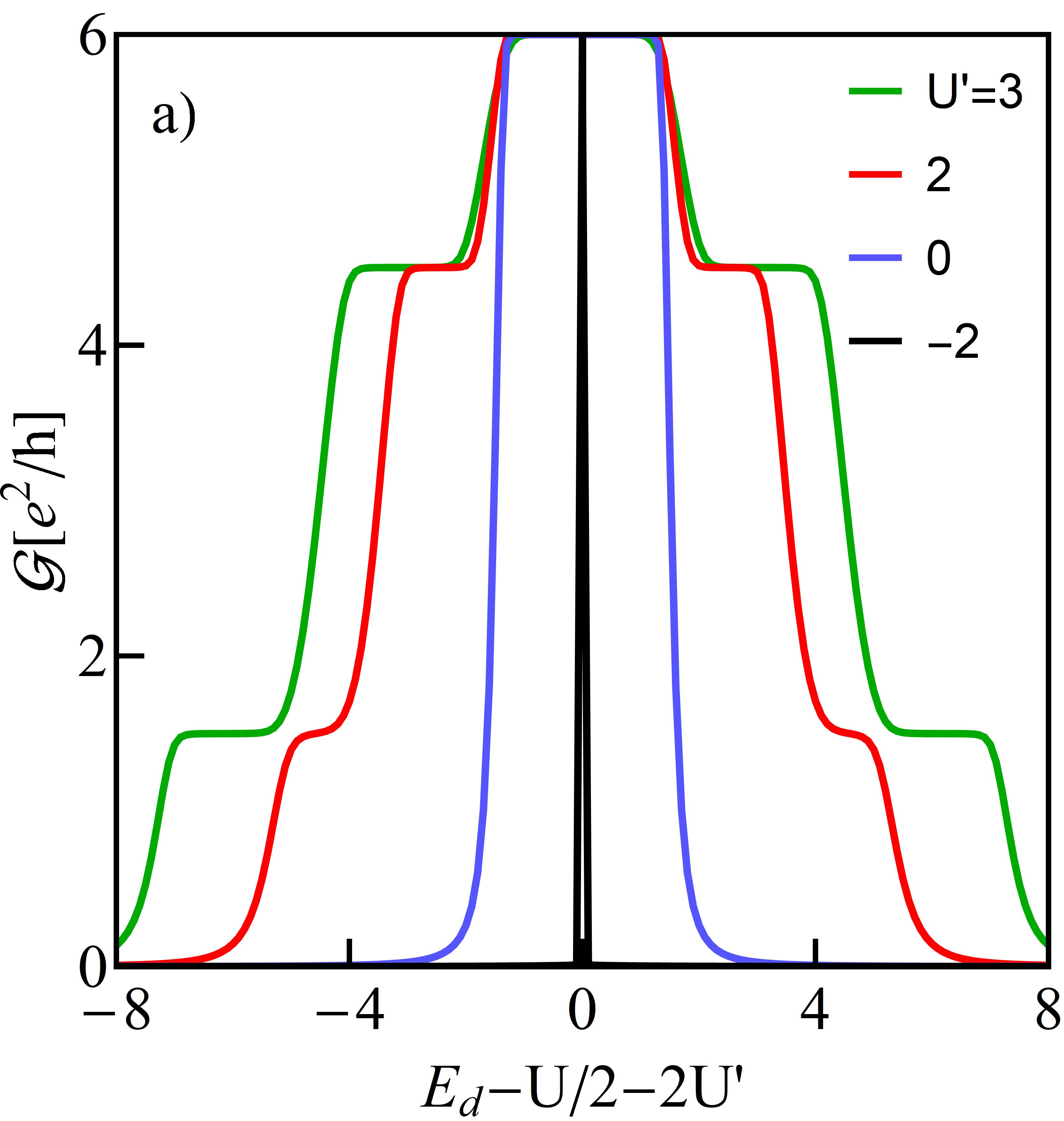}
\includegraphics[width=0.48\linewidth]{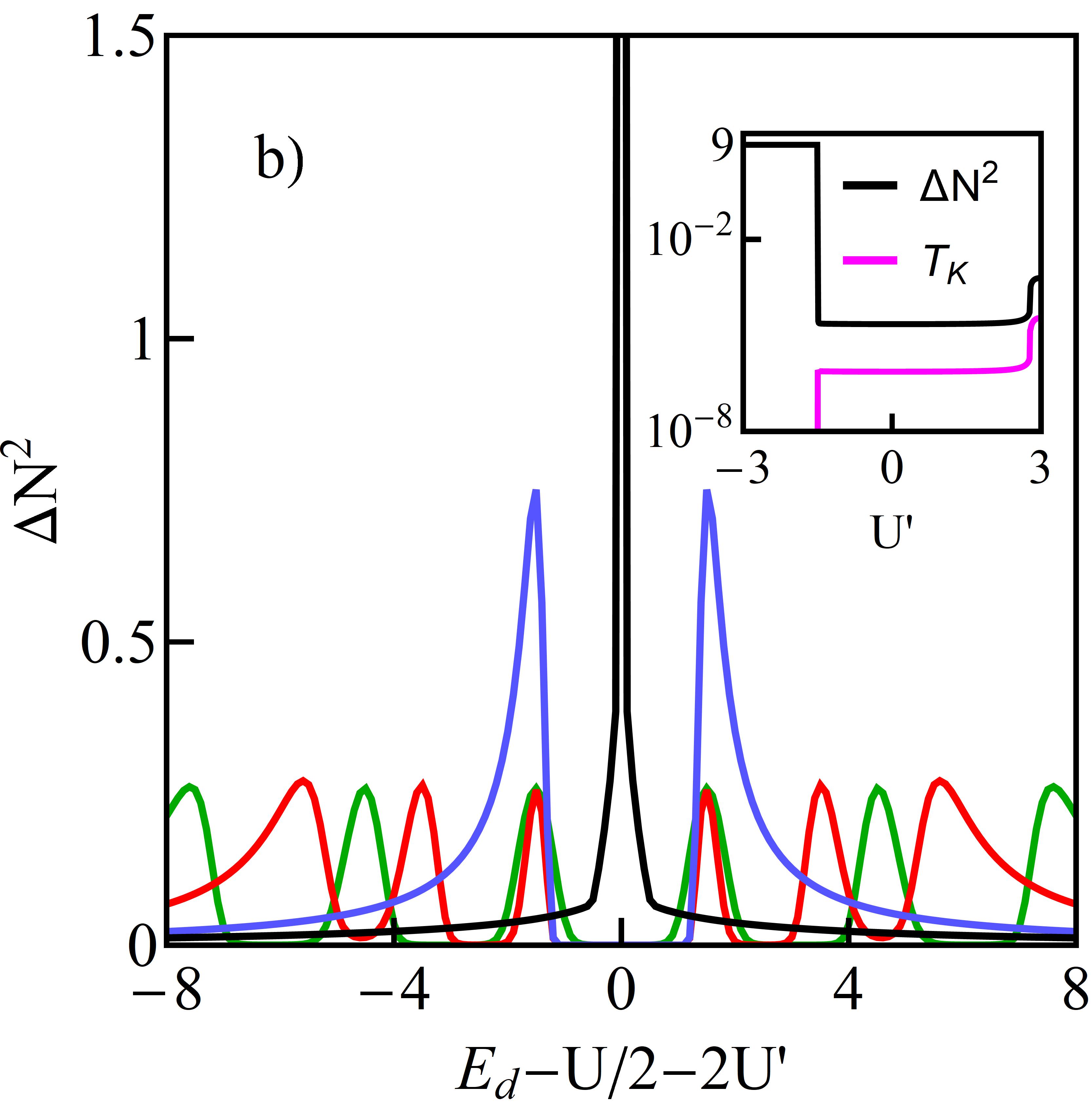}\\
\includegraphics[width=0.48\linewidth]{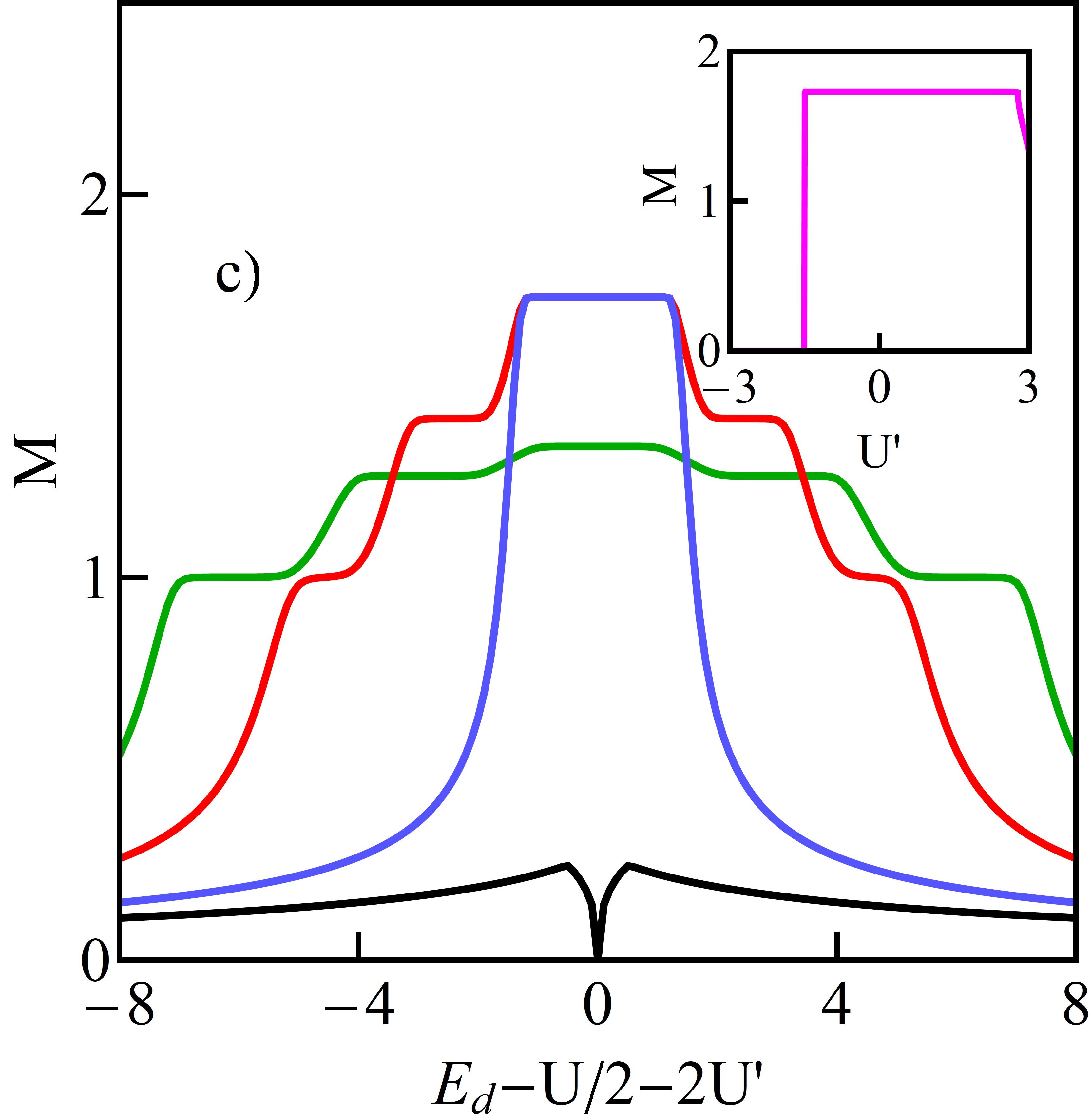}
\includegraphics[width=0.48\linewidth]{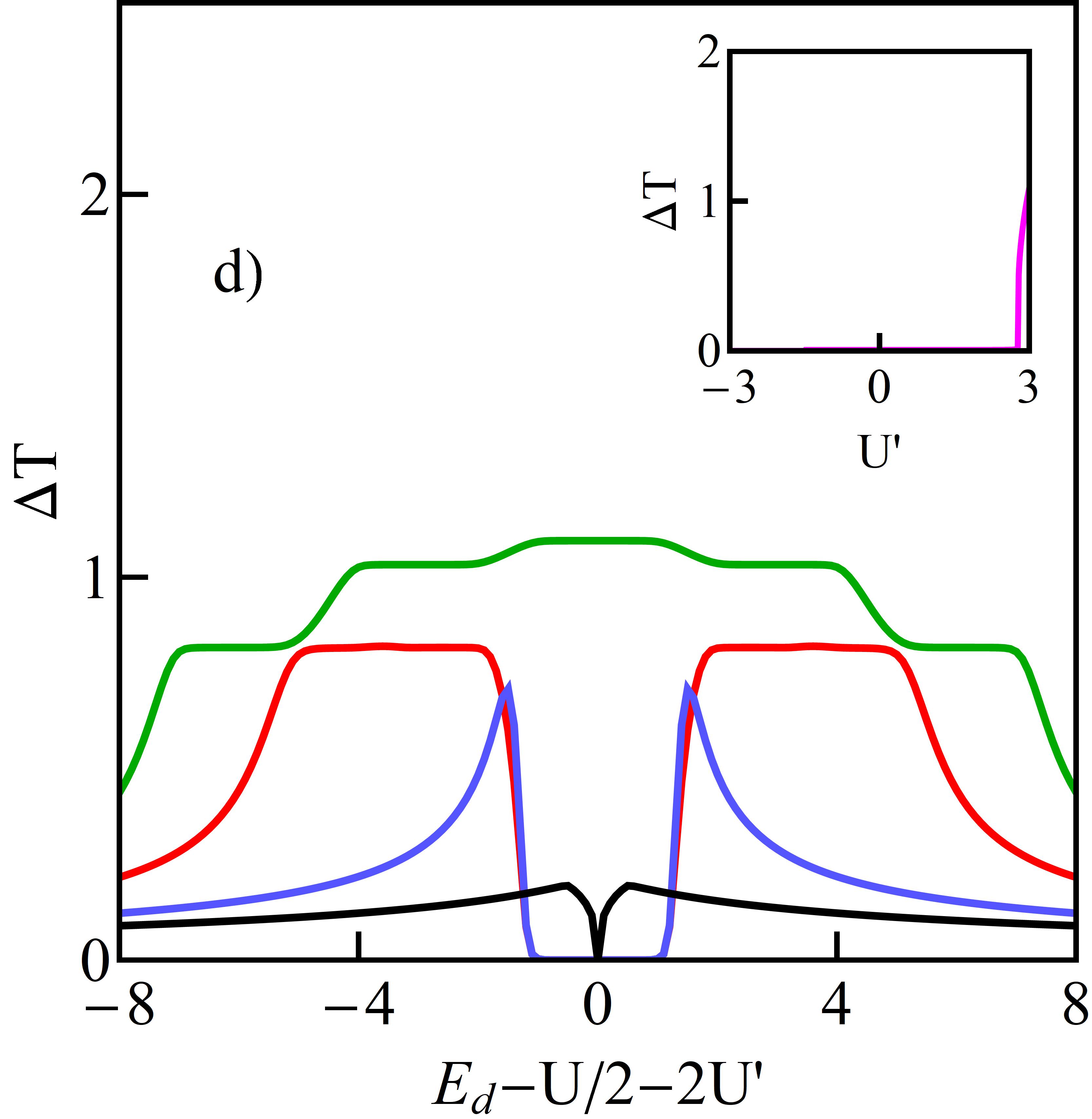}
\caption{\label{fig3} (Color online) a) Conductance for the selected interaction parameters $U'$,  $U' \leq U$  ($U = 3$)
b) Occupation fluctuations plotted for the same choice of interaction parameters as in Fig. 3a. Inset presents $\Delta N^{2}$ and Kondo temperature vs. interaction parameter $U'$ for e-h symmetry point. c),d) Local moment (c) and interdot occupation fluctuation (d) drawn for the same choice of interaction parameters as in Fig. 3a. Insets show dependencies of $M$ and $\Delta T$ on $U'$ respectively.}
\end{figure}
For $U' < U$  preferred is a separation of electrons   between different dots, which results in a large spin moment and a reduction in the differences in the occupancy of the dots. It is especially visible for $N = 3$, where $\Delta T$ goes to zero (Fig. 3b) and in equating $\Delta T$ values for $N = 1$ and $N = 2$ and  similarly for $N = 4,5$. For $U' < 0$  plateaus for $N = 1,2,4,5$ completely disappear and this for $N = 3$ successively narrows with decreasing $U'$ (Fig. 3c and Fig. 1b).
\begin{figure}[t!]
\includegraphics[width=0.48\linewidth]{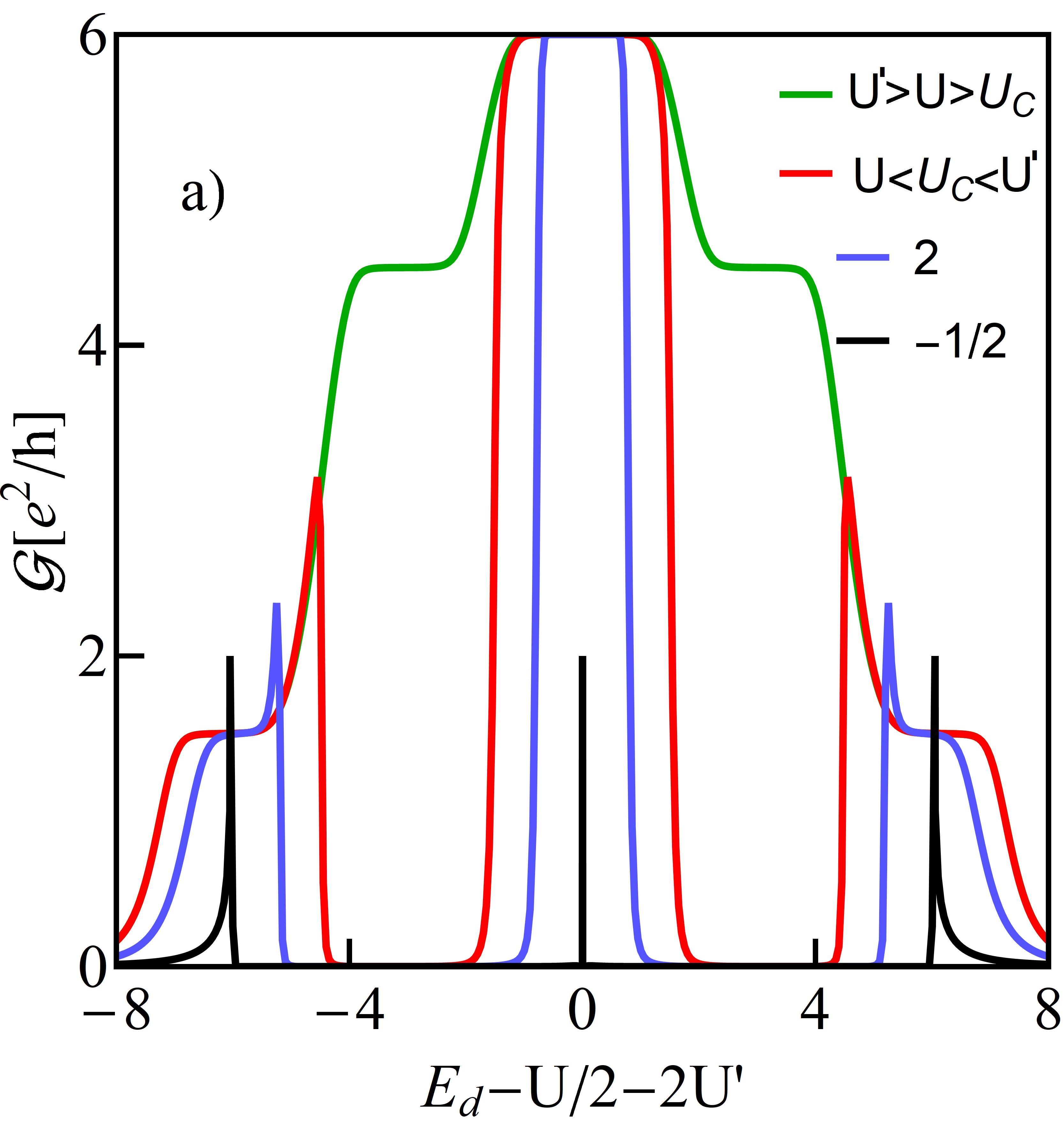}
\includegraphics[width=0.48\linewidth]{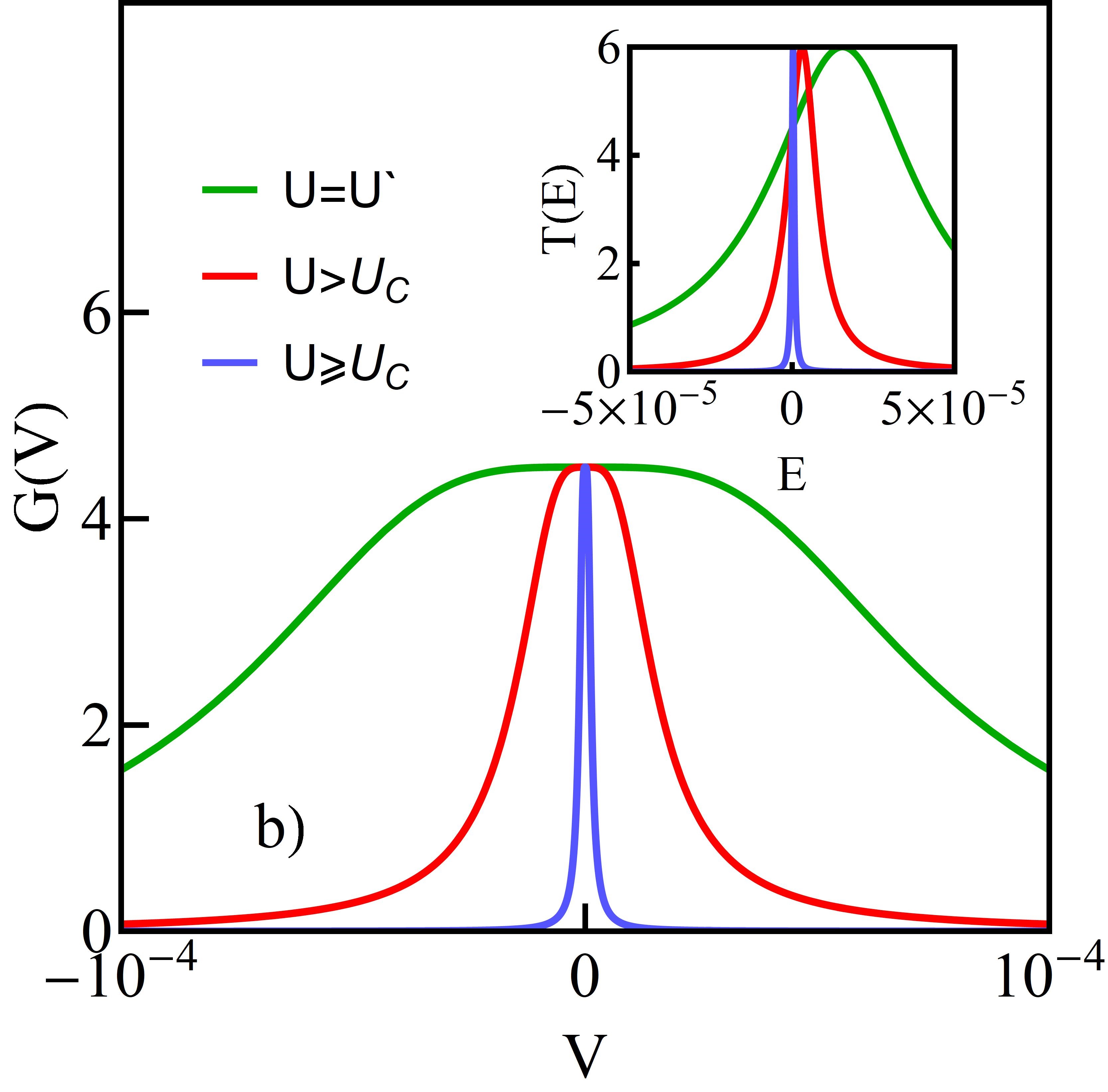}\\
\includegraphics[width=0.48\linewidth]{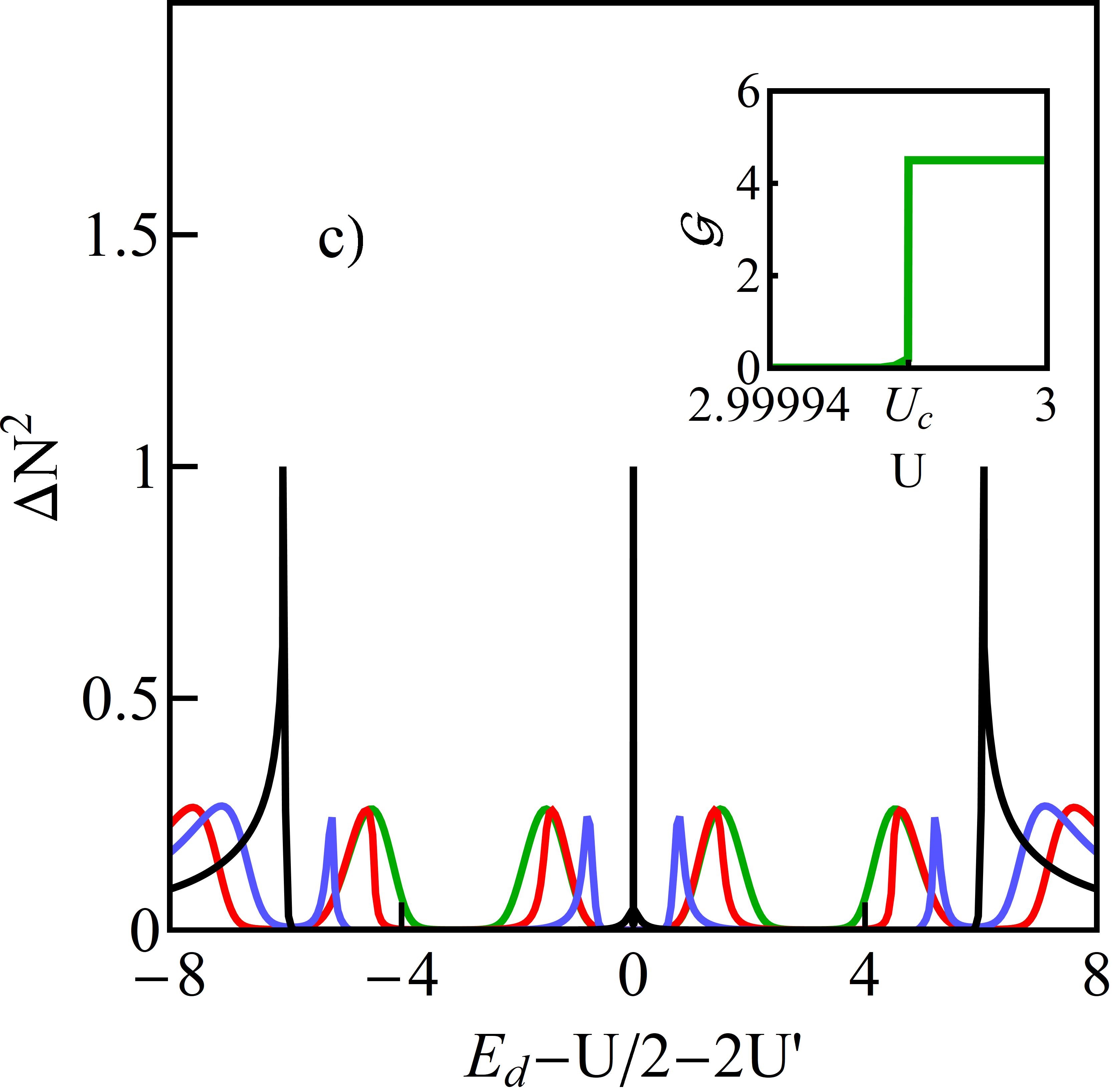}
\includegraphics[width=0.48\linewidth]{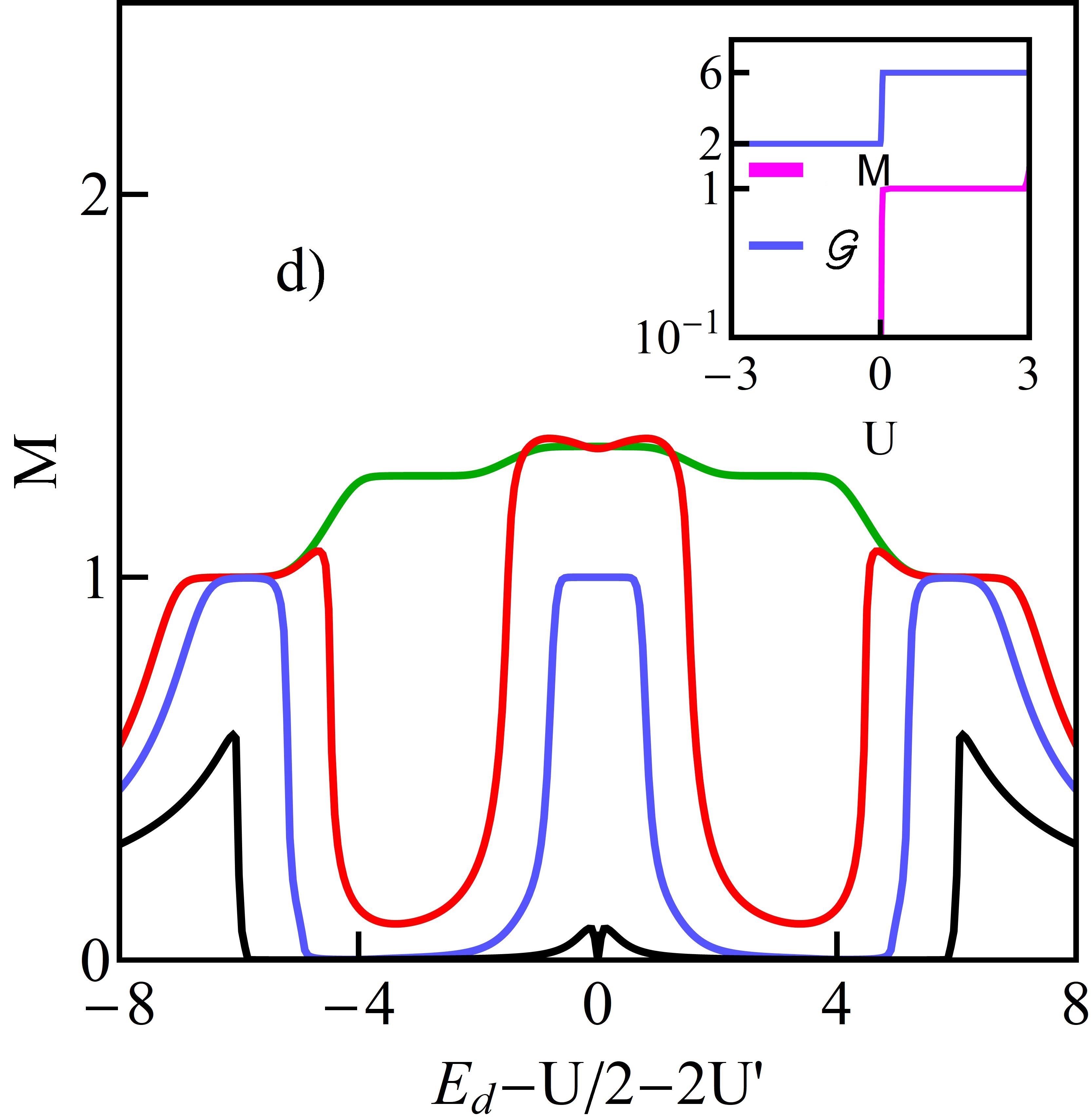}
\caption{\label{fig4} (Color online) a) Conductance for $U’ = 3$ and selected parameters $U$: $U = 2.9999$ ($U’>U>U_{c}=2.9997$) $U = 2.7$, $U =2$, and $U = -0.5$.   b) Low bias conductance for $N = 2$ and  $U = U’ = 3$, $U =2.9999$ ($U>U_{c}$)  and   $U = 2.9998$ ($U>U_{c}$) and in the inset transmissions corresponding to these parameters. c) Occupation fluctuations plotted for the same choice of interaction parameters as in Fig. 4a. Inset  illustrates vanishing of conductance with decreasing the intradot Coulomb interaction. d) Local moment  for interaction parameters from Fig. 4b and in the inset  local moment and conductance for e-h symmetry point.}
\end{figure}
For $U' = -U/2$ at the e-h symmetry point there occurs a degeneration between empty and fully occupied states. $e$ and $s$ bosons take the values  close to $1/\sqrt{2}$ and outside this point either $e \approx 1$ or $s \approx 1$ and the rest of boson amplitudes are small  (Fig. 3). For $U' \leq -U/2$ the effective $0 \rightarrow 6$ charge fluctuations (pseudospin fluctuations) lead to the formation of  charge SU(2) Kondo resonance reflecting in the occurrence of narrow conductance peak. It is the same type of resonance as mentioned earlier for $U = U'$ case ($U <  0$).

Figure 4a shows gate voltage dependence of conductance for different $U$ values, $U < U'$. For small difference of  $U'- U$ conductance is almost unaffected in odd occupation regions by weak symmetry breaking. However, significant changes in multi-body processes take place in  $N = 2$ and $N = 4$ manifolds already close to point $U = U'$. For  $U < U'$ in  $N = 2$ valley states with single occupancies of  the two dots are energetically higher than these with double occupancy of a single dot and when $U'-U$ exceeds Kondo temperature of SU(6) state, $U'-U>T^{SU(6)}_{K}$, one can expect quantum phase transition to the degenerate charge ordered  (CO) ground state  with $(2,0,0)$, $(0,2,0)$  or $(0,0,2)$ states for TQD. This is suggested by considering of  $U'\gg U$ limit and earlier calculations by numerical renormalization group method performed for similar system of two capacitively coupled dots \cite{Galpin}.
SBMFA formalism we use applied to  Hamiltonian (1) gives for $U>U_{c}$ solution $d_{1} = d_{2} = d_{3} = 1/\sqrt{3}$ and $z_{i\sigma} = 0$.   It correctly predicts transport properties, conductance drops to zero because the  dots become decoupled from the leads. However, the forseen ground state is not  CO state. The same problem arose in the NRG calculations of DQD and to get broken-symmetry charge ordered phase these authors supplemented the Hamiltonian with potential scattering term $H_{K} = K\sum_{i\sigma}(n_{i}-1)c^{\dagger}_{0i\sigma}c_{0i\sigma}$. We followed the same path and got solution $d_{i} = 1$ (e.g. $d_{1} = 1$) with rest of  bosons equal zero. Again all the dots are decoupled from the leads, but now CO phase  is predicted. Close to $U_{c}$, but  for $U > U_{c}$  effective tunneling between  $(2,0,0)$, $(0,2,0)$ and $(0,0,2)$ states  mediated by two-electron  states characterized by the single occupancy of the dots is still possible and these processes quench charge isospin (SU(3) charge  Kondo state with  $d_{1} = d_{2} = d_{3} = 1/\sqrt{3}$, small values of $d_{ij\sigma\sigma'}$ and completely negligible values of the rest of the bosons).  Analogous effect appears for  $N = 4$ with holes playing the same role as electrons for $N = 2$ (CK state with $f_{1} = 1$ and CO state with $f_{1} = f_{2} = f_{3} = 1/\sqrt{3}$). Fig. 4b  compares  low bias conductances and transmissions for  SU(6) Kondo state ($N = 1e$), for broken SU(6) resonance  and charge Kondo state. Transmission line narrows and shifts toward Fermi level and the corresponding Kondo temperature of CK resonance is extremely  small . As it is seen from charging diagram (Fig. 1d), for $U < 0$ direct transitions between empty and $N = 2$ CO state or completely filled ($N = 6$) and $N = 4$, CO state are possible. The degeneracy between two even-number charge states is the consequence of attractive interaction. Effective fluctuations between these states induce SU(2) charge Kondo effect ($e = d_{1}= 1/\sqrt{2}$ or $s = f_{1} = 1/\sqrt{2}$).\newline
  Summarizing, using slave boson approach we have considered a symmetrical, capacitively coupled system of three quantum dots in the strongly correlated regime for all occupations  and investigated the evolution of the system as a function of both the intradot and interdot coupling strengths. We have analyzed gate voltage dependencies of conductance, local magnetic moment, as well as fluctuations of these quantities. Due to a subtle interplay of spin and charge degrees of freedom the rich range of behavior is observed i.e. SU(6) spin- charge  Kondo effect, 3$\times$SU(2) spin Kondo effects, charge ordered states,  SU(3) charge Kondo effect preserving total charge of the system and charge Kondo effect with effective fluctuations between occupations $n = 0$ and $n = 6$.

\def\refname{References}

\end{document}